# Synthesis of calcium carbonate in trace water environments

Giulia Magnabosco,[a] Iryna Polishchuk,[b] Boaz Pokroy,[b] Rose Rosenberg,[c] Helmut Cölfen,[c] Giuseppe Falini*[a]

**Calcium carbonate ($CaCO_3$) was synthesized from diverse water-free alcohol solutions, resulting in the formation of vaterite and calcite precipitates, or stable particle supensions, with dimension and morphology depending upon the condition used. The obtained results shed light on the importance of water molecules during crystallization of $CaCO_3$ and open a novel synthetic route for its precipitation in organic solvents.**

The study of the $CaCO_3$ precipitation process is a key point for many different fields, from material science[1-3] to biomineralization.[4-6] The processes happening *in vivo* are of great significance even for developing new methodologies that can be applied *in vitro*.[7, 8] Although it is well known that organic molecules,[9-14] supersaturation, pH, template and temperature[15-18] play a fundamental role in the control of polymorphism, morphology and dimension of the crystals, even the hydration sphere of the involved ions is of crucial importance.[19, 20] In fact, water molecules influence strongly the reactions taking place during the $CaCO_3$ precipitation,[21] in particular the ones involved in the carbonate speciation. Few works have been done to investigate the role of the solvent during $CaCO_3$ crystallization, principally because of the difficulties in finding an appropriate solvent in which to perform the precipitation process. The ideal solvent must be able to dissolve salts and easily stay anhydrous. Among organic solvents, alcohols meet these requirements and, moreover, ethanol is the most used one in the studies present in the literature.[22-26]

When ethanol is present as an additive in an aqueous solution during a $CaCO_3$ precipitation process, it stabilizes vaterite and prevents its conversion to calcite.[26-28] In addition Sand *et al.*[29] showed how different alcohols, their concentration and the experimental parameters affect the stability, morphology and polymorphism of $CaCO_3$ in binary alcohol-water system, developing a model that is able to predict the outcome of the reaction based on the conditions used. When ethanol acts as a solvent, amorphous calcium carbonate (ACC) is the predominant polymorph obtained with the diffusion of ammonium carbonate into a solution of calcium chloride,[19] while using calcium hydroxide as starting material, a mixture of calcite, vaterite and aragonite is obtained.[24] In these reactions the formation of carbonate ions from diffusing gases (i.e. $NH_3$ and $CO_2$) implies the presence of water. To the best of our knowledge, no reports describing direct mixing of calcium and carbonate ions in an almost water free environment are present in the literature.

In this communication we describe a new simple method to precipitate calcium carbonate from alcohol solutions of anhydrous calcium chloride and ammonium carbonate. In this system the low quantity of water diminishes the rate of carbonate speciation, favoring the precipitation of only two products, $CaCO_3$ and ammonium chloride ($NH_4Cl$). The effect of different molecular weight (MW) alcohols, their volume ratio and the concentration of calcium and carbonate ions (as reported in Table 1) was investigated.

Anhydrous calcium chloride and ammonium carbonate were dissolved in absolute ethanol, then the solutions were added to absolute ethanol, or other alcohols, using syringe pumps under continuos magnetic stirring until reaching the desired concentration (see experimental section in ESI).

After 3 hours from the beginning of the reaction, only some samples produced $CaCO_3$ particles that could be separated by centrifugation at 4500 g (highlighted in grey in Table 1). The other solutions presented a suspension of $CaCO_3$.

No other reaction times were analysed, since the goal of this communication was to investigate the precipitation of $CaCO_3$ in the absence of water in diverse organic solvents, after a time when different behaviours were detectable. Shorter reaction time resulted in a general formation of suspensions and longer ones in the general formation of precipitates. In a further research the time-evolution of the $CaCO_3$ formation will be carried out.

The suspensions, which did not show a macroscopic precipitate were investigated by dynamic light scattering (DLS) and Analytical Ultracentrifugation (AUC). At 5 mM $CaCO_3$ concentration, nanoscopic species were detected by DLS for all

Table 2. Measurements of particles (nm) present in the solution still stable after a centrifugation at 4500 g for 10 minutes obtained using DLS. For the sample prepared using 5 mM salts in MeOH, it was not possible to measure particles since their concentration was too low.

| Solvent Conc. | MeOH | EtOH | 1-PropOH | 1-BuOH |
|---|---|---|---|---|
| 20 mM | 120.9 ± 0.73 | --- | --- | --- |
| 10 mM | 74.7 ± 0.1 | 65.7 ± 0.31 | --- | --- |
| 5 mM | ---$ | 35.7 ± 0.1 | 117.1 ± 0.32 | 22.0 ± 0.3 |

$ no measurable

Table 1. Table containing the concentrations and the solvents examined in this work. They were methanol (MeOH), ethanol (EtOH), 1-propanol (1-PropOH) and 1-butanol (1-BuOH). Samples prepared using 1-PropOH and 10 mM salts precipitated after 5 days (see experimental section in ESI† and Table SI1). The samples indicated in grey show precipitates while the white ones are stable dispersions.

| Solvent Conc. | MeOH | EtOH | 1-PropOH | 1-BuOH |
|---|---|---|---|---|
| 33 mM | 66% EtOH 33% MeOH | 100% EtOH | 66% EtOH 33% 1-PropOH | 66% EtOH 33% 1-BuOH |
| 20 mM | 40% EtOH 60% MeOH | 100% EtOH | 40% EtOH 60% 1-PropOH | 40% EtOH 60% 1-BuOH |
| 10 mM | 20% EtOH 80% MeOH | 100% EtOH | 20% EtOH 80% 1-PropOH | 20% EtOH 80% 1-BuOH |
| 5 mM | 10% EtOH 90% MeOH | 100% EtOH | 10% EtOH 90% 1-PropOH | 10% EtOH 90% 1-BuOH |

solvents (exception methanol), at 10 mM only for methanol and ethanol and at 20 mM only for methanol (Table 2). These data indicate that with increasing solvent polarity, nanoparticles can be stabilized against precipitation and that with decreasing $CaCO_3$ concentration, the particle size decreases. For 5 mM and 10 mM $CaCO_3$ in ethanol and 5 mM in 1-butanol, the particle size distributions could be determined by AUC, which were in good agreement with the DLS data and showed that the 5 mM samples were rather monodisperse (Fig. SI 1). However, the 5 mM and 20 mM $CaCO_3$ samples in methanol and 5 mM and 10 mM in ethanol as well as the 5 mM sample in 1-propanol contained very small species. Their sedimentation coefficients are shown in Fig. SI 2. The sedimentation coefficients are in the order of 0.1 – 0.3 S which is typical for ions / ion pairs with the exception of the 1-propanol sample.[30] A larger species is also detected with sedimentation coefficients of around 1S, which falls into the range of prenucleation clusters with the exception of the 1-propanol sample.[30] Partly, even larger species with sedimentation coefficients around 3 S are observed (Fig. SI 2). What these species are cannot be determined from these data. Therefore, we employed the diffusion coefficients, which can at least qualitatively be determined via fitting of the sedimentation raw data using the Lamm equation (Fig. SI 2). The particle diameters (d) could be calculated applying the Stokes-Einstein equation. With a modified Svedberg equation (1), the density

Table 3. Particle diameters (d) in nm and densities ($\rho$) in g/ml of small species detected via AUC. The subscripts 1-3 indicate families of particles having different diameter and density.

|  | d1 / $\rho$1 | d2 / $\rho$2 | d3 / $\rho$3 |
|---|---|---|---|
| MeOH 20 mM | 1.3 / 0.88 | 2.9 / 0.86 |  |
| EtOH 10 mM | 1.0 / 1.18 | 2.9 / 1.24 | 7.4 / 0.91 |
| EtOH 5 mM | 0.9 / 1.29 | 1.8 / 1.48 |  |

of the species can be estimated from the size and sedimentation coefficient as shown in Table 3.

$$\rho_i = \rho_0 + \frac{180\, \eta_0\, s}{d^2} \qquad \text{equation (1)}$$

where $\rho_i$ (in g/ml) is the density of the sedimenting particle, $\rho_0$ and $\eta_0$ (in Poise) are the density and the viscosity of solvent, s (in Svedberg) is the sedimentation coefficient and d (in nm) is the particle size (Stockes-equivalent sphere diameter).[31]

From Table 3, it can be seen that at least for methanol and ethanol, very small species could be detected with sizes around 1 nm for the smallest species 1, 2 – 3 nm for species 2 and 7.5 nm for species 3. The density of the species in methanol is markedly smaller than that in ethanol although the density of the solvents is almost equal (0.79 g/ml). This indicates a higher degree of solvatisation of the ionic calcium and carbonate species for the more polar methanol as compared to ethanol. The smallest detected species with a size of 0.9 nm - 1.3 nm could potentially be related to a solvated $CaCO_3$ ion pair (r $Ca^{2+}$ = 0.10 nm, r $CO_3^{2-}$ = 0.18 nm), while the larger species already must contain dozens of ions with a size similarity to prenucleation clusters for the water case.[30] The solubility of $CaCO_3$ in alcohol decreases with the increase of the alcohol MW, resulting in a higher $CaCO_3$ precipitation yield in high MW solvents even at lower starting salt concentration. The precipitates were characterized as collected after centrifugation and drying at 60 °C. No water washing was carried out to avoid any dissolution, and eventually, a re-precipitation process. The solid products were analysed by Fourier transform infrared (FTIR) spectroscopy, scanning electron microscopy (SEM) and synchrotron high resolution powder X-ray diffraction (HRPXRD). The FTIR and HRPXRD data showed that $CaCO_3$ co-precipitated with $NH_4Cl$, as a side product. The bands in the FTIR spectra 1475 $cm^{-1}$, at 876 $cm^{-1}$ and 746 $cm^{-1}$ (Fig. 1), correspond to $v_3$, $v_2$ and $v_4$, respectively, vibration modes of vaterite. The bands at 1420 $cm^{-1}$ at 712 $cm^{-1}$ indicate the presence of calcite traces in some precipitates, while that at 1403 $cm^{-1}$ indicates $NH_4Cl$. The presence of vaterite, $NH_4Cl$ and small amounts of calcite was also confirmed by the Rietveld analysis of the HRPXRD patterns (Fig. SI3, Table SI2). The co-presence of $NH_4Cl$ was laso confirmed by its sublimation after thermal treatment at 300 °C (Fig. SI 4). The Rietveld analysis also showed that 1-propanol was the best solvent for the precipitation of vaterite (>99 wt%). This may suggest that in this solvent the solubility of vaterite, and of potential prenucleation clusters (see AUC data) is lower with respect to other alcohols.

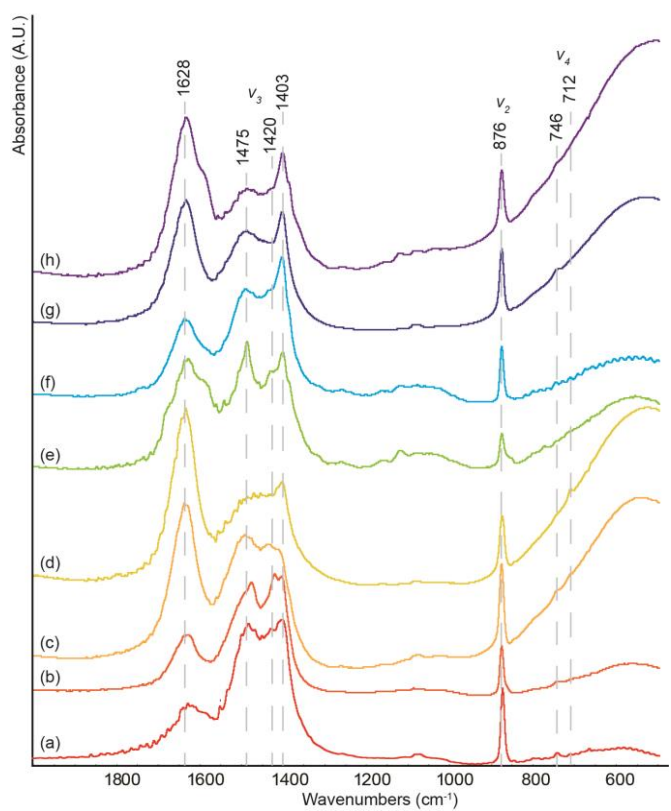

Figure 1 FTIR spectra of samples obtained using (a) MeOH and 33 mM salts, (b) EtOH and 33 mM salts, (c) EtOH and 20 mM salts, (d) 1-PropOH and 33 mM salts, (e) 1-PropOH and 20 mM salts, (f) 1-BuOH and 33 mM salts, (g) 1-BuOH and 20 mM salts and (h) 1-BuOH and 10 mM salts.

The intensities of the band at 3000 cm$^{-1}$ suggest a contribution of water to the ammonium absorption bands (Fig. SI 5). This indication is confirmed by the thermogravimetric analysis (Fig. SI 4), from which an amount of about 3-15 wt.% of linked water was detected in the precipitates, according to intensity of the absorption band at 1628 cm$^{-1}$.

Since after the precipitation process the quantity of water in the solution is lower than 0.5% (v/v), we can hypothesize that this water is collected from the environment due to its high affinity to the $CaCO_3$ surface and entrapped between the crystalline domains. Comparing all the samples prepared using 33 mM salt solutions it is possible to notice that the intensity of the band at 1628 cm$^{-1}$ increases with the MW of the selected alcohol. The lower solubility of water in alcohols with longer chain may promote the entrapping of water in the precipitate.

SEM imaging (Figure 2) reveals the presence of $CaCO_3$ particles with different morphologies together with an unstructured thin layer, probably of $NH_4Cl$ or amorphous calcium carbonate (not evident from HRPXRD and FTIR data) that covers the underlying material. The sample precipitated in the presence of methanol shows particles around 150 nm that assemble to form 3 μm aggregates with irregular shape. When the precipitation process is carried out using pure ethanol as solvent, particles with a more regular shape are present in the samples. Using 33 mM salt concentration, pillars with hexagonal section can be observed while, reducing the salt concentration to 20 mM, some elongated grain-like particles form. 1-propanol and 1-butanol precipitated samples seem to be influenced stronger by the concentration of the salts rather than by the nature of the solvent. In fact, for both samples prepared with 33 mM salts, it is not possible to recognize any regular shape. When decreasing the salt concentration to 20 mM and 10 mM, some spherical particles become visible due to the reduction of the covering layer. These particles are smaller than 1 μm and have an irregular surface (Fig. 2).

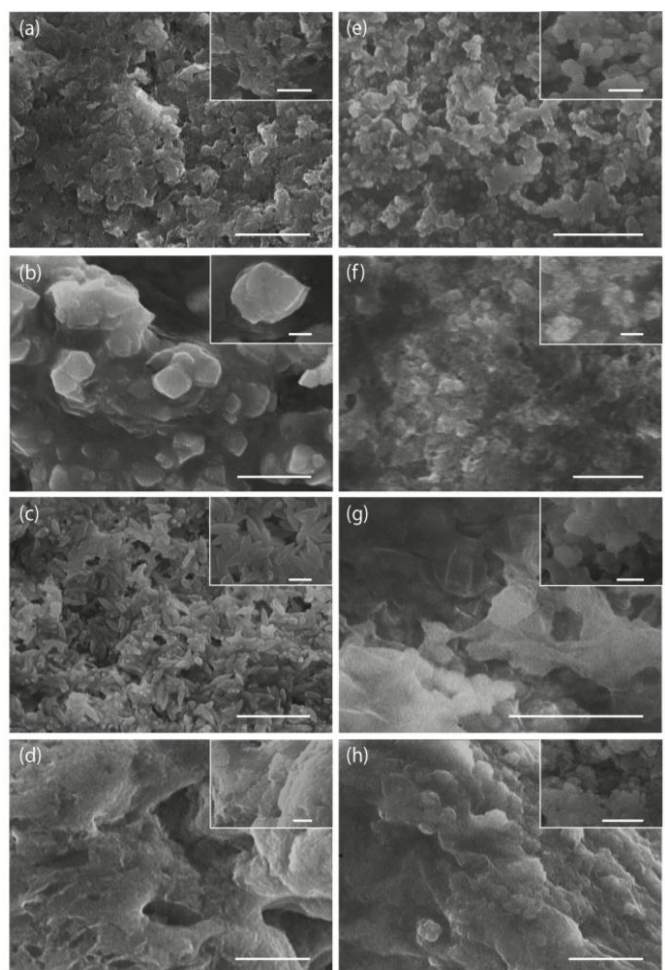

Figure 2 SEM images of samples obtained using (a) MeOH and 33 mM salts, (b) EtOH and 33 mM salts, (c) EtOH and 20 mM salts, (d) 1-PropOH and 33 mM salts, (e) 1-PropOH and 20 mM salts, (f) 1-BuOH and 33 mM salts, (g) 1-BuOH and 20 mM salts and (h) 1-BuOH and 10 mM salts. Scale bar is 5 μm in the main picture and 1 μm in the inset.

These results confirm that the solvent plays a fundamental role in the crystallization of $CaCO_3$ and add new information showing that the use of different alcohols stabilizes vaterite reducing its conversion to calcite, in agreement with previously published data.[30] The crystallization process in alcohol is slower than the one in water and, after 3 hours from ion addition, vaterite is the main component of the precipitate, while in pure water the same experimental conditions produce only pure rhombohedral calcite (Fig. SI 6 and SI 7). However, in similarity to the prenucleation clusters observed in water,[30] we could also detect several very small species in methanol and ethanol, which are likely solvated ions or their clusters.

In conclusion, this simple methodology can allow the study of the interaction between $CaCO_3$ and molecules that are not soluble in water, without the use of additional reactants, giving rise to new possible synthetic paths. Finally, the data show that the use of a different solvents significantly affects the $CaCO_3$ crystallization pathway, which will be the object of future further investigations.

We thank Dr. Davide Levy for helping in the analyses of the HRPXRD data. G.F. and G.M. thank Consorzio Interuniversitario di Ricerca sulla Chimica dei Metalli nei Sistemi Biologici for the support. B.P. thanks the European Research Council under the European Union's Seventh Framework Program (FP/2007-2013)/ERC Grant Agreement (no. 336077) for the support and the European Synchrotron Radiation Facility for the high resolution powder X-ray diffraction measurements at the beamline ID22.